# Reaction to New Security Threat Class

Yuval Elovici
Director of Telekom Innovation Laboratories
Head of Cyber Security Research Center
Department of Information Systems Engineering
Ben-Gurion University of the Negev, Israel
P.O.B. 653, Beer-Sheva, Israel 84105.
Fax: +972-8-6477527
email: elovici@bgu.ac.il

Lior Rokach
Department of Information Systems Engineering
Ben-Gurion University of the Negev, Israel
P.O.B. 653, Beer-Sheva, Israel 84105.
Fax: +972-8-6477527
email: liorrk@bgu.ac.il

**Abstract**

Each new identified security threat class triggers new research and development efforts by the scientific and professional communities. In this study, we investigate the rate at which the scientific and professional communities react to new identified threat classes as it is reflected in the number of patents, scientific articles and professional publications over a long period of time. The following threat classes were studied: Phishing; SQL Injection; BotNet; Distributed Denial of Service; and Advanced Persistent Threat. Our findings suggest that in most cases it takes a year for the scientific community and more than two years for industry to react to a new threat class with patents. Since new products follow patents, it is reasonable to expect that there will be a window of approximately two to three years in which no effective product is available to cope with the new threat class.

# 1. Introduction

Armies around the world are commonly criticized for their perceived lack of preparedness - during wartime, they find themselves fully prepared for the previous war fought, and they find themselves completely surprised and unprepared for coping with their opponents' latest weapons and strategies introduced in the current conflict.

In this paper we investigate how the professional and scientific security communities react to new security threat classes. We define a new threat class as a well-defined group of threats that have similar properties such as goal, method of performing the attack, etc Does the scientific community tend to provide timely solutions to potential threats as they emerge or alternatively, does it provide solutions to threats only once these begin to inflict damage? While some may argue that providing a solution to a hypothetical threat is problematic, as there is no way to evaluate the performance of the solution objectively against a collection of existing threats. Others might argue that putting a preemptive solution in place may prevent potential threat classes from ever emerging or inflicting any significant damage. In this study we look at the speed with which the scientific and the professional communities react to a newly defined threat class. Of course a threat class may exist long before it was defined and named by the scientific and professional community, and it is important to note that this represents one of this study's limitations.

We carried out this study by examining how fast the professional and scientific community reacts to a new emerging threat class. To quantify the response we investigated the relationship between the emergence of a new threat class and the number of associated publications and patents over time. Because the number of threat classes is enormous, we limited our study to a few threat classes with unique names commonly used by the professional and scientific communities. Their unique name is useful in identifying the relevant publications and in identifying the first time the threat class was mentioned.

The following threat classes were used:

- Phishing (Reid, 2009)
- SQL Injection (Strom, 2011)
- BotNet (Kola, 2008)
- Distributed Denial of Service (DDoS) (Mirkovic and Reiher, 2004)
- Advanced Persistent Threat (APT) (Websense, 2011)

We began by identifying when the new threat class initially emerged by determining the first time it was mentioned in a relevant context. We then totaled the number of professional articles, scientific papers and patents mentioning the threat class over time. Some may point out that using a new buzzword to define a threat class does not mean that the threat and the efforts to confront them did not previously exist. For example, the AIDS disease existed long before it was detected and named. We are aware of this potential criticism, yet we believe that, as in the case of the AIDS disease, when the number of victims increases, the threat is more thoroughly identified and more formally labeled. Following the definition of a new threat class it is easier for the community to develop dedicated mitigation techniques.

In the remainder of the paper we introduce the threat classes with a description of each threat class, including the date of its first reference and a brief description of the incident that brought it to the attention of the professional and scientific communities. We also provide information sources for each threat class and describe where we derived information concerning the threat class and list the number of professional articles, scientific publications and patents associated with each threat class over time. The paper ends with a final summary and conclusions.

## 2. Threat Classes

The following five threat classes were investigated in this study:

### 2.1 Phishing

Phishing is considered one of the most common social engineering attacks. In a phishing attack the adversary's goal is to obtain the victim's personal information and/or security credentials by using fraudulent email messages (Hong, 2012). These messages are composed in such a way that the innocent victim is made to believe that they stem from a legitimate source. The information obtained in a phishing attack could include personal information which may be used for identity theft, and in many cases, tends to include account numbers, user names and passwords. Because phishing attacks rely on innocent internet users, as well as the fact that nowadays messages are sent from BotNets, it is very difficult to stop these attacks completely (Purkait, 2012).

According to Reid (2009), the term 'phishing' originates from adversaries using e-mails to 'fish' for passwords and other private data from innocent Internet users. It is speculated that the letters, 'ph', are connected to naming conventions used by adversaries such as 'Phreaks'. Phishing was first mentioned in 1996, in reference to adversaries who were stealing America Online (AOL) accounts. The first mention of the term phishing on the Internet was in "2600 hacker newsgroup" in January 1996, though the hacker community might have used this term earlier. The earliest public media reference to phishing was in March 1997. Tatiana Gau, Vice President of Integrity Assurance for AOL stated that, "The scam was called 'phishing' as in fishing for your password, but spelled differently." In 1997, Ed Stansel, reporter for the Florida Times Union newspaper, was quoted as saying, "Don't get caught by online 'phishers' angling for account information." In this study, we will assume that the first reported phishing incident was in 1996 and that the first media report occurred in 1997.

## 2.2 SQL Injection

SQL Injection is an attack in which the adversary exploits the security vulnerabilities in a poorly designed web application (Clarke, 2012). These vulnerabilities allow an attacker to insert an SQL statement in the data entry field of the web application. The web application thereafter combines the inserted SQL statement with the statement that processes the data entry field. The consequence of the attack is that the attacker may read, update, or delete data stored in the database of the web application. Customarily, the goal of the attacker is to extract confidential information stored in the database and is generally associated with an intention to expose this stolen material via the web application. The confidential information may include credit card numbers and social security numbers. SQL Injection is an attack that can be easily prevented by the proper design of web applications. Unfortunately, today SQL Injection is still considered to be one of the most common attacks on web applications.

The first SQL injection incident was reported in 1996 (Puppy, 1998). It raised concerns that in many implementations the user input to queries is not checked and is assumed to be compliant with the expected input.

## 2.3 BotNet

A BotNet is a group of compromised computers which are under the control of one attacker (Rodríguez-Gómez et al., 2013). Each compromised computer is considered to be a "robot", thus deriving the term BotNet (roBOT NETwork). The victim's computer is compromised by the installation of a Trojan horse which is in contact with the attacker via a command and control channel. Usually the BotNets are designed in such a way that the users of the victim's computer do not suffer from a significant performance degradation of their computers. BotNets are commonly used to launch other types of security attacks such as sending SPAM messages, distributing malware and launching distributed denial of service attacks. In many cases BotNets are created by one group of attackers which specializes in this domain and then are "rented or sold" to other attackers in order to perform their attacks.

During the late 1990s, it was reported that several worms exploited the vulnerabilities in IRC clients such that the clients were remotely controlled (for example, IRC/Jobbo). SubSeven Trojan version 2.1 appeared in 1999 and allowed the SubSeven server to be remotely controlled by a bot connected to an IRC server. This development inspired the development of all the malicious BotNets. Thus, in this study we will assume that the first BotNet was reported in 1999.

**2.4 Distributed Denial of Service (DDoS)**

Attacks focusing on preventing the legitimate user from receiving a service are called Denial of Service (DoS) attacks. When the attacker uses multiple attacking entities in order to conduct the attack, the attack is called Distributed Denial of Service attack (DDoS). DDoS attacks may be launched by an attacker via a BotNet. DDoS attacks are characterized by the means to prepare and launch the attack, the properties of the attack and its effect on the attacked system/service (Mirkovic and Reiher, 2004). It is not easy to cope with DDoS attacks since it is very difficult for the defense mechanism to differentiate between malicious and benign traffic.

DDoS attack was first mentioned by SANS Institute in August 1999. In this attack, 200 compromised computers attacked a computer at the University of Minnesota (flooding attack). Thus, in this study we will assume that the first reported DDoS attack incident was in 1999.

**2.5 Advanced Persistent Threat**

Advanced Persistent Threats (APTs) are considered to be a new emerging threat; however, APTs very likely existed for many years before the term was used to describe a specific group of attacks. APTs are "cyber-attacks mounted by organizational teams that have deep resources, advanced penetration skills, specific target profiles and are remarkably persistent in their efforts" (Tankard, 2011). APT attacks are commonly launched by governments with great resources and usually exploit several vulnerabilities in order to achieve their goals. The goal of an APT is almost always very specific, such as disabling a particular function of the attacked entity or stealing specific information. APTs tend to work stealthily below the detection radar and are persistent in their activity until they achieve their goal.

It is widely accepted by the computer security community that APTs were first mentioned by the U.S. Air Force, circa 2006, in order to describe complex (i.e.,

"advanced") cyber attacks against specific targets over long periods of time (i.e., "persistent"). Thus in this study, we shall assume that the first reported APT incident was in 2006. It is known that APT attacks existed before 2006, but at that time they were not distinguished from other attacks.

## 3. Bibliographic Databases and Tools

To measure research and development efforts by the scientific and professional communities for each of the five threat classes (as reflected by mediums including relevant papers and patents), we searched various bibliographic databases, each of which covers different materials or subjects (see Table 1).

**Table 1:** Bibliographic Database

| Bibliographic Database | Content Coverage |
|---|---|
| Computer Database INFOTRACK | An index for news and reviews in the computer domain. |
| EBSCO Business Source | Covers academic papers, market research reports, industry reports and mass media. |
| Engineering Village | Includes the following bibliographic databases: Compendex (scientific and technical engineering research), Inspec (computer science, information technology, etc.) and EI Patents (US and European Patents). |
| Google Scholar | Provides a search of scholarly literature across many disciplines and sources. |
| IEEE Xplore | Index for all IEEE journals and conferences. |
| ISI Web of Science (Thomson Reuters) | A reference and citation database that publishes the ISI impact factor for journals. |
| Lexis-Nexis | An index for business and legal news. |
| ProQuest | A multi-disciplinary index for academic papers, professional articles and general news (e.g. the New York Times). |
| ScienceDirect (Elsevier) | Provides a search and full text access to journals and books. |
| Scopus (Elsevier) | An abstract and citation database of research literature and reliable web sources. |
| SpringerLink | Index for Springer journals, books and conferences (including lecture notes in computer science). |

We extracted the bibliographic records of all papers written from 1990 to 2012 that match one of the attack queries i.e., Advanced Persistent Threat (or APT), BotNet, Phishing, SQL Injection, and Distributed Denial of Service (or DDOS) in May 2013. Table 2 specifies the number of papers extracted from each Bibliographic Database by Publication Type (scientific papers, professional articles and patents).

**Table 2:** Number of Papers extracted from each Bibliographic Database

| Bibliographic Database | Number of Papers by Type | | |
|---|---|---|---|
| | **Academic** | **Professional and News** | **Patents** |
| EBSCO Business Source | 514 | 2202 | |
| Engineering Village | 6418 | 1283 | 9 |
| Google Scholar | 45839 | 3276 | 7160 |
| IEEE Xplore | 1765 | | |
| INFOTRACK | 170 | 6620 | |
| ISI Web of Science | 1972 | | |
| Lexis-Nexis | | 24086 | |
| ProQuest | 1685 | 23137 | 6 |
| ScienceDirect (Elsevier) | 4210 | | |
| Scopus (Elsevier) | 7790 | | |
| SpringerLink | 8140 | | |
| **Total** | **78503** | **60604** | **7175** |

We used RefWorks to consolidate records from all sources. As the same reference can appear in more than one source, we used the "duplicate" feature in RefWorks to identify redundant references and subsequently merged the duplicates to create a single entry. Contradicting fields are resolved by counting the occurrences of each value and choosing the most common one. If values have equal frequencies, the value that was obtained from the source that is deemed to be more accurate is selected. For example, if the same paper appears both in Google Scholar and in Thomson Reuters Web of Science, but these sources happen to disagree on a certain field then the value of Thomson Reuters Web of Science is chosen for this field. The following priority list is used for prioritizing the sources: IEEE Xplore --> SpringerLink --> ScienceDirect --> ISI Web of Science --> Scopus --> ProQuest --> Lexis-Nexis --> Computer Database INFOTRACK --> Engineering Village --> EBSCO Business Source --> Google Scholar. In total the meta-data of 146,282 papers have been extracted. After completing the deduplication process 72,221 items were left (about 49%).

Our data analysis (described below) requires that each document be classified as a single medium (academic, professional or patent). While most documents can easily be classified based on their publication outlet (an article published in Dr. Dobb's Journal will most likely be referred to as professional) or source (a paper indexed by SpringerLink is assumed to be academic), approximately 8% of the papers were automatically classified using ensemble learning algorithms (Rokach and Maimon, 2001; Menahem et al., 2009). We assumed that mediums are mutually exclusive, i.e., the same paper cannot be associated with more than one medium.

## 4. Results

In order to determine how quickly the scientific and professional communities and industry responded to a new threat class, we tracked significant milestones for each threat class by determining when each threat class was first mentioned in different mediums. Milestones included when the threat was first reported or defined, when the first patent was filed, when the first scientific paper was published mentioning the threat and when the threat was first mentioned in a professional article. The data for all threat classes is summarized below in Table 3.

**Table 3**: Milestones - date threat class was mentioned in different mediums

| Threat class | First reported (year) | Paper Associated with the first report | First mentioned in professional article | First scientific publication | First patent application |
|---|---|---|---|---|---|
| APT | 2006 | (Websense, 2011) | 2008 | 2008 | 2008 |
| BotNet | 1999 | (Canavan, 2005) | 1999 | 2000 | 2004 |
| DDoS | 1999 | (Preimesberger, 2013) | 1999 | 2000 | 2001 |
| Phishing | 1987 | (Felix and Hauck, 1987) ( Reid, 2009) | 1988 | 1988 | 2004 |
| SQL Injection | 1998 | (Puppy, 1998) | 1998 | 1998 | 2004 |

One can quickly observe in Table 3 that in most cases it took a year for the scientific community to respond to a new threat class. It is reasonable to assume that patents reflect industry reaction to a new threat class (i.e., industry development of new products associated with the threat), and our study shows that it took industry between one and six years to react to a new threat class. One possible explanation for industry's delayed reaction may be based on the need for industry to see a solid market demand related to a new threat class.

Following the initial analysis of the material presented in presented in Table 3, we also determined the number of references in publications from 1998 through 2012 for each of the threat classes. The results are summarized below (Figures 1-5).

In nearly all threat classes the number of references in professional publications is higher than that of scientific publications (with the exception of SQL Injection). It is interesting to note as well, that on average there is a patent for each 10 scientific publications (with the exception of APT).

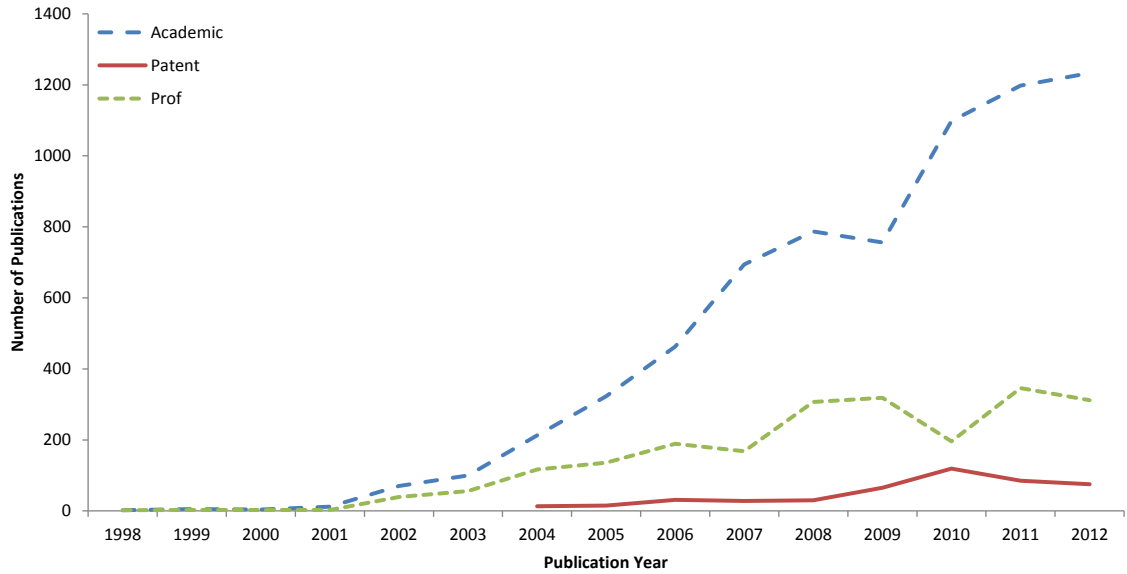

**Figure 1:** Publications related to SQL Injection

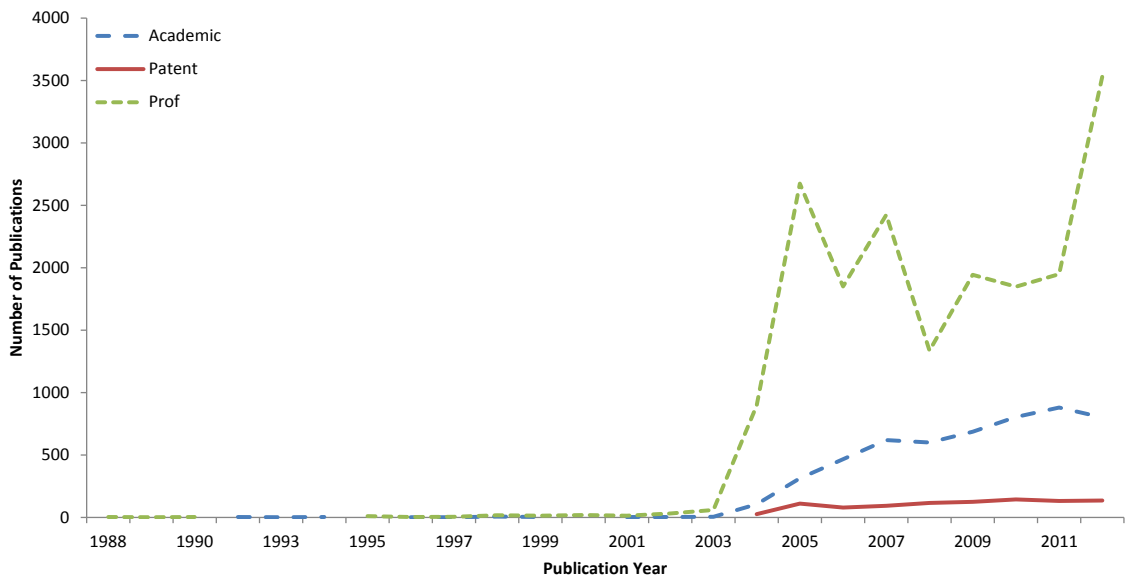

**Figure 2:** Publications related to Phishing

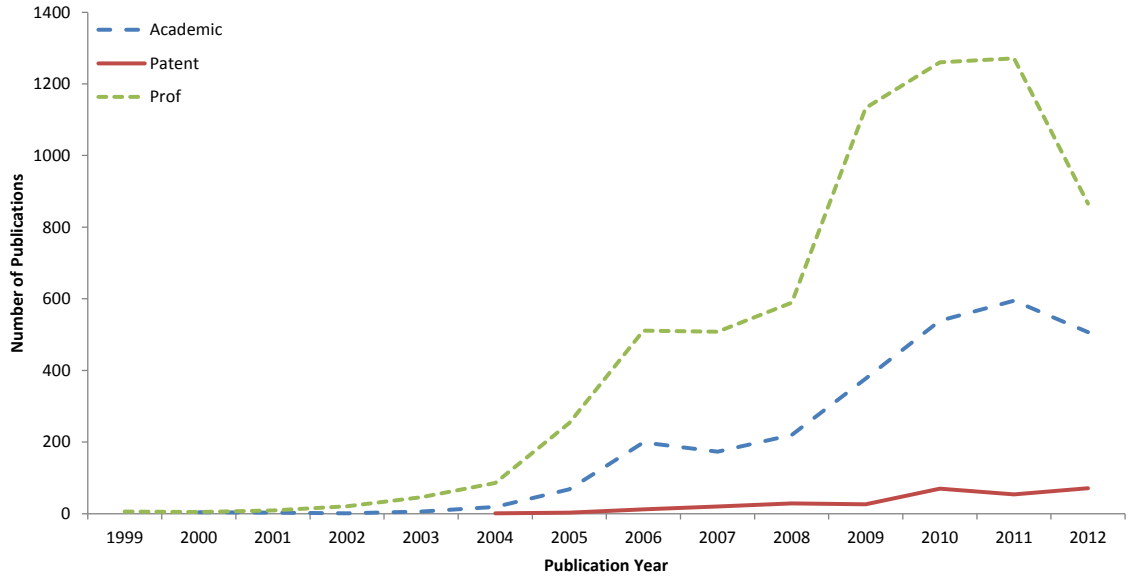

**Figure 3:** Publications related to BotNet

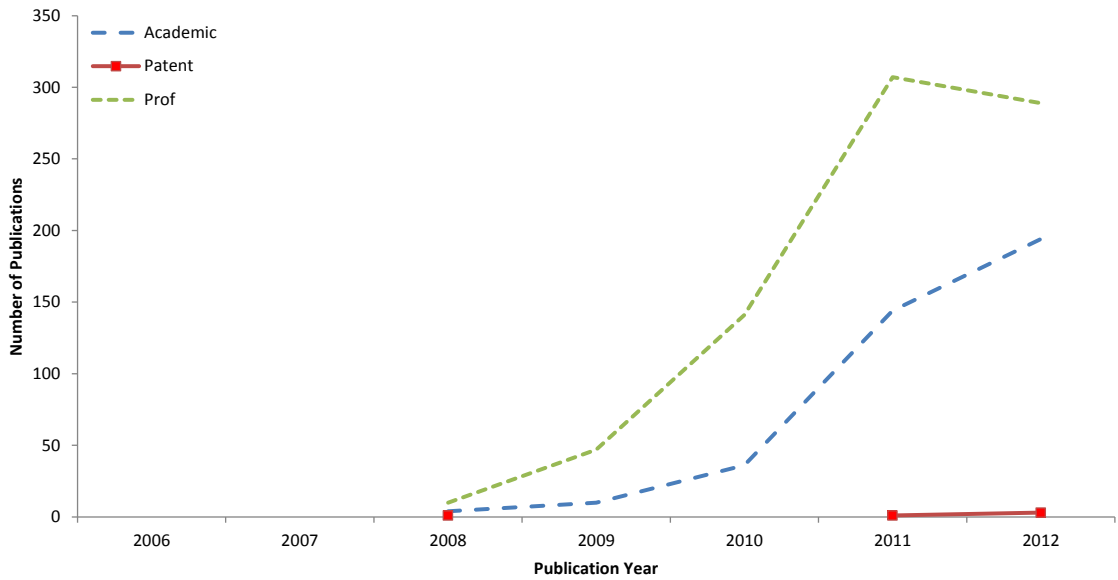

**Figure 4:** Publications related to APT

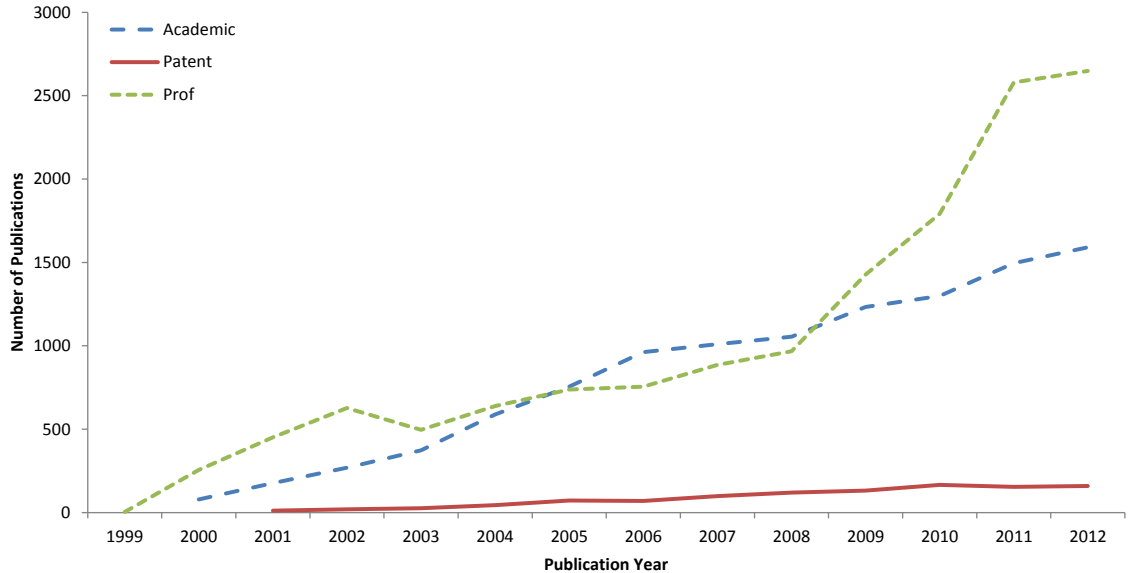

**Figure 5:** Publications related to DDoS

## 5. Summary and Conclusions

Each year, new security threat classes are identified and defined, mainly by the professional community. Each of the defined threat classes includes a group of threats with similar goals and each is based on the same attack methods. The definition of a new threat class helps the professional and scientific communities share information concerning newly developed mitigation technologies.

Our study shows that the scientific community reacts faster to a newly defined threat class compared to the professional community and industry (as reflected by the number of scientific publications, professional articles and patents). One possible explanation might be that the professional community and industry need further evidence that there is new market potential for a new class of products. It is reasonable to assume that it will take at least one year from the application date of the first patent to the appearance of a new product based on this patent. As a result, one can assume that systems may not be well protected against a newly defined threat class for at least two to three years from the time the threat class was first defined. This window of opportunity may be exploited by attackers who may react faster than the scientific community, the professional community or industry to new opportunities.

One of our study's primary limitations is that it does not present the correlation between the number of publications, articles and patents related to a threat class with the amount of damage inflicted by incidents associated with the threat class. Such an analysis may shed some light on whether the scientific and professional communities and industry are striving to mitigate relevant attacks. This important analysis was not performed, because we were unable to find reliable sources for the number of incidents associated with each threat class and damage estimates for each incident.